\documentclass[aps,prb,showpacs,twocolumn,groupedaddress]{revtex4}
\usepackage[dvips]{graphicx}
\usepackage{ulem}
\usepackage{hyperref}

\def\be{\begin{equation}}
\def\ee{\end{equation}}
\def\vec{\mathbf}
\def\mc{\mathcal}

\begin{document}
\title{Microscopic Theory for the Markovian Decay of Magnetization Fluctuations in Nanomagnets}
\author{Ioannis Rousochatzakis} 
\email{ioannis.rousochatzakis@epfl.ch}
\affiliation{Institut de th\'eorie des ph\'enom\`enes physiques, 
Ecole Polytechnique F\'ed\'erale de Lausanne,\\
CTMC, BSP - Dorigny, CH-1015 Lausanne, Switzerland}

\begin{abstract}       
We present a microscopic theory for the phonon-driven decay of the magnetization fluctuations in a wide 
class of nanomagnets where the dominant energy is set by isotropic exchange and/or uniaxial anisotropy. 
Based on the Zwanzig-Mori projection formalism, the theory reveals that the magnetization fluctuations 
are governed by a single decay rate $\omega_c$, which we further identify with the zero-frequency 
portion of the associated self-energy. This dynamical decoupling from the remaining slow degrees of freedom 
is attributed to a conservation law and the discreteness of the energy spectrum, and explains the 
omnipresent mono-exponential decay of the magnetization over several decades in time,  
as observed experimentally. A physically transparent \textit{analytical} expression for $\omega_c$ 
is derived which highlights the three specific mechanisms of the slowing down effect which are known so far 
in nanomagnets.
\end{abstract}


\maketitle

\section{Introduction}
Over the past two decades there has been an intense theoretical and experimental interest in the field of
molecular nanomagnets~\cite{Kahn,GSV} in view of their importance in the general context of quantum 
magnetism but also for potential applications, e.g. in quantum computing~\cite{Appl_QC,Tejada},
memory storage~\cite{Appl_MS}, and magnetic imaging~\cite{Appl_MI}. 
The dominant energy in these ``zero-dimensional'' magnets is set by intramolecular isotropic exchange
interactions and/or easy-axis anisotropy. 
Two of the most central motifs in this research field have been\cite{GSV} 
the fact that the decay of the \textit{equilibrium} fluctuations $J(t)$ 
of the total magnetization $S_z$ is characterized by a single correlation frequency parameter 
$\omega_c$ (or inverse decay time $1/\tau_c$) \textit{over several decades in time}, 
and the experimental manifestation of a dramatic slowing down of $\omega_c$ 
which establishes already from relatively high temperatures $T$.  

The correlation frequency parameter is a quantity of broad experimental relevance 
and of great theoretical value. For instance, according to linear response theory 
and the fluctuation-dissipation theorem, $\omega_c$ is probed by ac (longitudinal) susceptibility 
and magnetization relaxation measurements. 
It may also be measured directly by Electron Spin Resonance~\cite{exp3} 
since $\omega_c$ is the electronic spin-lattice relaxation rate $(1/T_1)_{\text{el}}$.
Moreover, if $\omega_c$ is in the regime of the nuclear Larmor frequency $\omega_L$, 
then it can be indirectly probed by Nuclear Magnetic Resonance (NMR) as in Antiferromagnetic 
Rings~\cite{NMR_Baek,NMR_Santini,NMR_Borsa}. 
Similarly, $\omega_c$ is also of relevance in Muon~\cite{exp1} and M\"{o}ssbauer~\cite{exp2} spectroscopies. 
Previous theoretical studies of $\omega_c$~\cite{Villain_MEs,GSV,Loss_MEs,Wurger,Bessais} 
are based on the Master Equation (ME) which embodies the coupled dynamics of the full set of 
populations $p_l$ of the various eigenstates $|l\rangle$ of the isolated spin Hamiltonian $\mc{H}_s$.
In this approach, one first evaluates the various phonon-driven quantum-mechanical transition 
rates $W_{ll'}$ from $|l\rangle$ to $|l'\rangle$ and then forms and diagonalizes 
numerically the associated relaxation matrix $\mc{R}$. 
This procedure gives a multi-exponential decay.  
To account for the observed behavior it is then sometimes argued~\cite{GSV,Loss_MEs,Wurger,Bessais} 
that $\omega_c$ can be assigned to the lowest eigenvalue (or decay rate) $\lambda_0$ of $\mc{R}$. 
Despite being physically quite plausible, such a hypothesis alone does not guarantee that $\lambda_0$ 
corresponds to the decay of the observable of interest (here $S_z$), since other observables 
(such as the magnetic energy $\mc{H}_s$) decay slowly as well. 
As shown by Santini \textit{et al.}~\cite{NMR_Santini}, a full numerical treatment of this problem 
based on the ME approach requires also the calculation of the weight of each eigenvalue of $\mc{R}$ 
in $J(t)$. In agreement with experimental data, the ME calculations of 
Santini \textit{et al.}~\cite{NMR_Santini} showed that $J(t)$ is dominated by a single eigenvalue 
and which, indeed, does not always equal $\lambda_0$. A principal theoretical question however still remains 
as to whether or not there is a physical reason which necessitates the observed dynamical decoupling of $S_z$ 
over several decades in time and in a large class of magnetic molecule clusters. 

In this article we put forward a theoretical framework which focuses \textit{right from the start} on the 
principal quantity of interest, namely the fluctuations $J(t)$ of the total magnetization $S_z$.  
This approach reveals immediately the physical origin of the observed Markovian decay:
The self-energy $\Sigma(t)$ of the magnetization embodies fluctuations of non-conserved observables 
which decay much faster than $S_z$. It is shown that this separation of time scales is guaranteed 
by the discreteness of the magnetic energy spectrum.
The identification of the precise eigenvalue of $\mc{R}$ which governs the decay of $J(t)$ 
with the zero-frequency portion of the self-energy and the derivation of an analytical 
expression for it (cf. Eq.~(\ref{omegac1.Eq}) or (\ref{omegac2.Eq})) in lowest order in the 
magnetoelastic coupling is a second achievement of this approach. 
This expression is an optimal starting point for direct and quantitative comparison with 
experimental data and thus allows to explore the specific low-energy phonon modes 
which are involved in the magnetoelastic process in all nanomagnets, 
an issue which is of broad interest and is currently largely unexplored. 
At the same time, this expression readily demonstrates the three specific mechanisms 
of the slowing down effect which are known so far in nanomagnets: 
(i) the existence of an easy-axis anisotropy, (ii) the increasingly gapped character 
of the spectrum at low energies, and (iii) the possibility of a crossover to a critical slowing down. 

The first is present in Single Molecule Magnets (SMM's)~\cite{GSV,Sessoli_SMMs,Sang_SMMs} which 
are perhaps the most intensively studied nanomagnets since, apart from the above ``super-paramagnetic'' 
behavior, they also provide a remarkable manifestation of magnetization tunneling~\cite{Chudnovsky_tunn}. 
Here, the presence of competing exchange interactions stabilizes a ferrimagnetic ground state 
of a large total spin $S$, e.g. $S=10$ for the Mn$_{12}$ and Fe$_8$ clusters. 
The energy scale within this large $S$ manifold is set by an effective uniaxial anisotropy $-|D|S_z^2$, 
giving rise to a correspondingly large barrier 
($U\simeq 62$ K for Mn$_{12}$ and $24.5$ K for Fe$_8$)~\cite{GSV,Sessoli_SMMs,Sang_SMMs} for thermal relaxation. 
The relaxation time $\tau_c$ of $S_z$ follows a phenomenological Arrhenius law $\tau_c \propto e^{U/k_BT}$ 
which can reach values as long as $10^5-10^7$ sec at low $T$. 
This behavior is also revealed by complementary ac susceptibility measurements~\cite{GSV}. 

The second mechanism appears e.g. in Antiferromagnetic Rings AFMR's. 
These are planar ring-like magnetic clusters comprising an even number $N$ of spins $s$
whereby the dominant energy is set by the nearest-neighbor antiferromagnetic (AFM) exchange $J$. 
This stabilizes a non-magnetic $S=0$ ground state separated by a spin gap 
$\Delta \simeq 4 J/N$ (Land\'e interval rule) from the lowest $S=1$ excitation. 
Here there is no energy barrier and thus the corresponding $\omega_c$ are on a much faster scale than the ones 
in SMM's. Still, it has been found~\cite{NMR_Baek,NMR_Santini,NMR_Borsa} that $\omega_c$ drops by several 
(three up to five) decades as $T$ drops below $\Delta/k_B$. This dramatic decrease is systematically revealed by an enhancement 
of the nuclear spin-lattice relaxation rate $1/T_1$ 
at $T \sim \Delta/k_B$~\cite{NMR_Baek,NMR_Santini,NMR_Borsa}. 
A similar behavior of $1/T_1$ 
is observed in other AFM clusters but also in SMM's~\cite{NMR_Belesi}. 
In the latter, the analysis of $1/T_1$ is hindered by the NMR wipe-out effect~\cite{NMR_Belesi}. 

The third mechanism of the slowing down effect emerges in 
Single Chain Magnets (SCM's)~\cite{SCM1,SCM2,SCM3,SCM4,SCM5} 
which comprise finite ferromagnetic chains of variant lengths $L$. Here, a crossover from an activated 
to a critical slowing down occurs as we tend toward the $T=0$ (thermodynamically) critical point, 
and in particular when the correlation length $\xi$ approaches $L$. Our approach accommodates such a 
mechanism exactly as in critical bulk systems~\cite{Forster} since it reveals that $\omega_c$ is inversely 
proportional to the isothermal magnetic susceptibility $\chi_0$ (cf. Eq.~(\ref{omegac1.Eq})).

The remainder of this article is organized as follows. In Sec.~\ref{Ham.Sec} we describe the main terms 
of the spin Hamiltonian of the magnetic cluster and its interaction with the host lattice which drives the 
relaxational dynamics. In Sec.~\ref{Mori.Subsec} we introduce the so-called Mori correlation function 
and express the main quantities of interest in terms of it. 
The Zwanzig-Mori projection technique is then employed in Sec.~\ref{Self.Subsec} 
in order to introduce the self-energy function and highlight its major role. 
A perturbative expansion of this quantity  is then given up to lowest (second) order 
in the magnetoelastic coupling.
In Sec.~\ref{Pert.Subsec} we provide a rigorous justification of the Markov approximation 
based on the discreteness of the magnetic energy spectrum. We then arrive at our expression 
for the correlation frequency and provide a qualitative analysis of the three specific 
mechanisms of the slowing-down effect which are known so far in nanomagnets. 
In Sec.~\ref{NMR.Sec} we revisit the problem of nuclear spin-lattice relaxation in AFMR's 
and clarify a number of aspects in the light of our treatment. 
Finally we give a short summary and overview of our results in Sec.~\ref{Summary.Sec}.

\section{Hamiltonian}\label{Ham.Sec}
At not too low $T$, and given the typical energy scale of the magnetic excitations 
(a few tens of degrees Kelvin), 
the decay of the electronic spin fluctuations must originate from the coupling 
to the deformations of the host lattice, and in particular, the \textit{low-energy} phonon modes.  
The Hamiltonian of the full spin$+$phonon system can be generally written as 
\begin{equation}
\mc{H}=\mc{H}_s+\mc{H}_{ph}+\mc{V}~,
\end{equation}
where $\mc{H}_s$ stands for the isolated spin Hamiltonian, $\mc{H}_{ph}$ is the phonon energy and 
$\mc{V}$ denotes the relevant spin-phonon interaction channel(s). 
The first is taken to include the most dominant magnetic energy terms  
\be\label{Hs.Eq}
\mc{H}_s = \sum_{ij} J_{ij} \vec{s}_i\cdot \vec{s}_j -|D| S_z^2 + g \mu_B B S_z ~,
\ee
i.e., the isotropic Heisenberg exchange interactions (leading term), 
a possible effective easy-axis anisotropy (second term), and finally a Zeeman term for the coupling 
between an external longitudinal (along the easy $z$-axis) field $\vec{B}$ and the total 
magnetic moment $S_z=\sum_i s_i^z$. As usually, $g$ denotes the electronic
spectroscopic factor and $\mu_B$ is the Bohr magneton.  In what follows, we shall denote by $|l\rangle$ and 
$E_l$ the eigenstates and eigenvalues of $\mc{H}_s$ and by $M_l$ the corresponding 
eigenvalues of $S_z$. 
In our treatment we have excluded Dzyaloshinskii-Moriya interactions and 
transverse on-site anisotropies which are very small ($\lesssim 1$ K) and are relevant  
at ultra low temperatures only. 
As we show below, the observed mono-exponential decay of $S_z$ can be explained by the most dominant 
energy terms which are included in Eq.~(\ref{Hs.Eq}). 

For the magnetoelastic coupling $\mc{V}$, we may write without loss of generality 
\begin{equation}
\mc{V} = \sum_\mu \mc{A}^\mu\otimes\mc{B}^\mu~, 
\label{coupling}
\end{equation}
where $\mc{A}^\mu$ and $\mc{B}^\mu$ are spin and phonon operators respectively. 
The latter represent low-energy strain fields probed by the electronic spin degrees of freedom. 
Details about the origin and possible types of $\mc{V}$ are discussed e.g. 
in Refs.~\onlinecite{GSV,Villain_MEs,Garanin,Abragam,Luthi}. 
As we show below, the major phonon quantities that govern the dissipative dynamics of the 
electronic spin fluctuations are the spectral densities  
\be
J^{ph}_{\mu\mu'}(\omega)\equiv\int_{-\infty}^{\infty} dt~e^{i\omega t}
\big\langle\mc{B}^\mu\mc{B}^{\mu'}(t)\big\rangle_{\text{ph}}~, 
\ee
where the thermal average is taken over the canonical ensemble $\rho_{ph}(T)=e^{-\beta \mc{H}_{ph}}/Z_{ph}$, 
$Z_{ph} =$ Tr $e^{-\beta\mc{H}_{ph}}$.

\section{Magnetization fluctuations}
In the absence of spin-phonon coupling, $S_z$ is conserved since $[S_z,\mc{H}_s]=0$. 
If the observables coupled to $S_z$ (through $\mc{V}$) do not share this property
it is physically expected that they decay faster than $S_z$.   
This separation of time scales is at the origin of the Markovian decay of $S_z$. 
And such a central physical ingredient can be exploited naturally within the framework of Zwanzig-Mori 
projection formalism~\cite{Mori,Zwanzig,Fick,Zubarev,Kubo,Forster}, 
whereby one may isolate the important slow degrees of freedom by appropriately projecting out the remaining 
faster ones. To this end, it is expedient to first express the central quantities of interest 
in terms of the so-called Mori correlation function which can be readily handled with the Zwanzig-Mori 
projection technique.

\subsection{Quantities of central interest in terms of the Mori correlation function}\label{Mori.Subsec}
Let us first introduce the Mori correlation function. 
We first note that any operator $A$ can be represented as a ``state vector'' $\big|A\big)$ 
in the Liouville space (space of operators). A convenient choice of a scalar product in this vector space 
is the so-called Mori product defined as 
\begin{eqnarray}\label{MoriProduct.Eq}
\big(A\big | B \big) \equiv \frac{1}{\beta}\int_0^\beta dx 
\big\langle A^\dagger B(i\hbar x)\big\rangle
=\big\langle A^\dagger\frac{1-e^{-\beta\hbar\mc{L}}}{\beta\hbar\mc{L}}B\big\rangle,
\end{eqnarray}
where $B(i\hbar x)\equiv e^{-\hbar\mc{L}x} B = e^{-x\mc{H}} B e^{x\mc{H}}$, and  
$\mc{L}=\mc{L}_s+\mc{L}_{ph}+\mc{L}_{\mc{V}}\equiv \mc{L}_0+\mc{L}_\mc{V}$  
denotes the Liouville operator. Its action on a given operator $A$ is defined as 
\be
i\mc{L}A \equiv \frac{1}{i\hbar}[A,\mc{H}]=\frac{d A(t)}{d t}\Bigg|_{t=0}\equiv \dot A~.
\ee
Also, the thermal average is taken over the canonical ensemble $\rho (T)$ of
the full spin $+$ phonon system
\be\label{rho1.Eq}
\rho (T) = Z^{-1} e^{-\beta \mc{H}},~Z=\text{Tr } e^{-\beta \mc{H}}
\ee 
where $\beta=1/(k_BT)$ denotes the inverse temperature and $k_B$ is the Bolzmann's constant.
In what follows we assume that $\beta\mc{V} \ll 1$ and thus the magnetoelastic coupling 
cannot influence static properties, i.e. the canonical density matrix factorizes as 
\be\label{rho2.Eq}
\rho(T) \simeq \rho_s (T) \otimes \rho_{ph}(T)
\ee 
where $\rho_{ph}(T)$ was defined previously while $\rho_s(T)=e^{-\beta \mc{H}_s}/Z_s$, 
$Z_s=$Tr$e^{-\beta \mc{H}_s}$.
Now, the Mori correlation function $M(t)$ is nothing else but the projection of the time-evolved ``state'' 
$\big|\delta S_z(t)\big)=e^{i\mc{L}t} \big|\delta S_z\big)$ onto $\big|\delta S_z\big)$, i.e.  
\begin{eqnarray}\label{Mt.Eq}
M(t)\equiv\big(\delta S_z \big| \delta S_z(t)\big) 
=\Big\langle \delta S_z \frac{1-e^{-\beta\hbar\mc{L}}}{\beta\hbar\mc{L}} \delta S_z(t)\Big\rangle~, 
\end{eqnarray}
where $\delta S_z \equiv S_z-\langle S_z\rangle$ and
$\delta S_z(t)\equiv e^{i \mc{H} t/\hbar} \delta S_z e^{-i \mc{H} t /\hbar}=e^{i \mc{L} t} \delta S_z$.  
The one-sided Fourier (or Laplace) transform of $M(t)$ reads   
\begin{eqnarray}\label{Ms.Eq}
\mc{M}(s)&\equiv&\big(\delta S_z \big|\delta S_z\big)_s = 
\int_0^{\infty}dt~e^{i s t}~\big(\delta S_z\big|e^{i\mc{L}t}\delta S_z\big)\nonumber\\ 
&=&\big(\delta S_z\big|\frac{i}{s+\mc{L}}\delta S_z\big)~.
\end{eqnarray}

On the other hand, the experimentally most relevant quantities which embody the dynamics of the 
magnetization fluctuations are the spectral density $\mc{J}(\omega)$ and 
the dynamical susceptibility $\chi(s)$. The first is the Fourier transform of the product correlation 
function $J(t)$, i.e.
\be\label{spectral.Eq}
\mc{J}(\omega)\equiv\int_{-\infty}^{\infty} dt~e^{i\omega t} \big\langle\delta S_z\delta S_z(t)\big\rangle~,
\ee 
while the second is the one-sided Fourier transform of the commutator correlation function 
\be\label{chi1.Eq}
\chi(s)\equiv\int_0^{\infty}dt~e^{i s t} \big\langle\frac{1}{i\hbar}[\delta S_z,\delta S_z(t)]\big\rangle,
~s=\omega+i\eta,~\eta>0.
\ee
The two quantities are connected via the fluctuation dissipation theorem
\be\label{fluctdiss.Eq}
\mc{J}(\omega) = \frac{2\hbar}{1-e^{-\beta \hbar \omega}} \chi''(\omega+i 0^+)~.
\ee
It is now straightforward to express the dynamical susceptibility $\chi(s)$ 
in terms of $\mc{M}(s)$ using their definitions Eqs.~(\ref{Ms.Eq}) and (\ref{chi1.Eq}), and the Kubo identity 
$\beta \big(A\big|\dot B\big)=-\big\langle \frac{1}{i\hbar}[A^\dagger,B]~\big\rangle$ which follows 
from Eq.~(\ref{MoriProduct.Eq}) by replacing $B$ with $\dot{B}$ and using the relation
$\big\langle A^\dagger e^{-\beta\hbar\mc{L}} B\big\rangle=\big\langle B A^\dagger\big\rangle$. 
The resulting expression is  
\begin{eqnarray}\label{chi2.Eq}
\chi(s) &=& -\beta\big(\delta S_z\big|\delta\dot S_z\big)_s = 
\beta\big(\delta S_z\big|\frac{\mc{L}}{s+\mc{L}}\delta S_z\big)\nonumber\\
&=& \chi_0 + i \beta s~\mc{M}(s)~,
\end{eqnarray}
where $\chi_0$ is the isothermal differential susceptibility: 
\begin{eqnarray}\label{chi0.Eq}
\chi_0 &\equiv& \beta \big(\delta S_z\big|\delta S_z\big)
=\beta\Big\langle\delta S_z\frac{1-e^{-\beta\hbar\mc{L}}}{\beta\hbar\mc{L}}\delta S_z\Big\rangle\nonumber\\
&=&\beta\big\langle(\delta S_z)^2\big\rangle=-\frac{1}{g\mu_B}\frac{\partial\langle S_z\rangle}{\partial B}~.
\end{eqnarray}
The second line follows from $\beta\mc{L}\simeq\beta(\mc{L}_s+\mc{L}_{ph})$ 
(since $\beta \mc{V} \ll 1$) and $\hbar L_s S_z =[\mc{H}_s,S_z]=0$.

\subsection{Projection method and self-energy $\Sigma(s)$}\label{Self.Subsec}
Having at hand Eqs.~(\ref{chi2.Eq}) and (\ref{fluctdiss.Eq}), we may now concentrate on the 
Mori correlation function which can be easily handled within the Zwanzig-Mori formalism. 
The idea is to project out from Eq.~(\ref{Ms.Eq}) fluctuations that couple to $S_z$ through 
the resolvent $\mc{G}=i (s+\mc{L})^{-1}$. This can be done by employing the projection operators 
\be
\mc{P}=\frac{\big|\delta S_z\big)\big(\delta S_z\big|}{\big(\delta S_z\big|\delta S_z\big)}~,~ 
\mc{Q}=1-\mc{P}~,
\ee
and then by decomposing $\mc{G}$ using the operator identity     
$(\mc{C}+\mc{D})^{-1}=\mc{C}^{-1}-(\mc{C}+\mc{D})^{-1}\mc{D}\mc{C}^{-1}$, 
with $\mc{C}=s+\mc{Q}\mc{L}$, and $\mc{D}=\mc{P}\mc{L}$. After some straightforward steps we obtain  
\be\label{MsSelf.Eq}
\mc{M}(s) = \frac{\chi_0}{\beta} \frac{i}{s+\Omega + i\Sigma(s)}~,
\ee
where we have used $M(t=0)=\big(\delta S_z \big| \delta S_z\big)=\chi_0/\beta$, and  
introduced the eigenfrequency $\Omega$ and the self-energy or memory function $\Sigma(s)$ 
which are defined by 
\begin{eqnarray}
i\Omega &=&\frac{\beta}{\chi_0} \big(\delta S_z \big| i \mc{L}\delta S_z\big) = 
\frac{\beta}{\chi_0} \big(S_z \big| \dot S_z\big) \label{Omega.Eq}\\ 
\Sigma(s)&=& \frac{\beta}{\chi_0}\big(\delta S_z\big|\mc{L}\mc{Q}
\frac{i}{s+\mc{Q}\mc{L}\mc{Q}}\mc{Q}\mc{L}\big|\delta S_z\big)\label{Self.Eq}~.
\end{eqnarray}
The physical meaning of these quantities becomes transparent by taking the inverse Laplace 
transform of Eq.~(\ref{MsSelf.Eq}) which reads
\be\label{DiffEq.Eq}
\frac{d}{dt} M(t) = i \Omega M(t) - \int_0^t dt' \Sigma(t') M(t-t'),~t\geq 0. 
\ee
This justifies the origin of the terms eigenfrequency and memory function for $\Omega$ 
and $\Sigma$ respectively.  

Equations~(\ref{MsSelf.Eq}), (\ref{Omega.Eq}) and (\ref{Self.Eq}) are exact. 
All information about the remaining fluctuations coupled to $S_z$ has been 
conveniently embedded into $\Sigma(s)$.  
Here $\Omega =0$ due to the definite time reversal signature of $S_z$ (also due to the Kubo
identity given above). Importantly, this also implies that 
$\mc{Q}\big|\dot S_z\big)=(1-\mc{P})\big|\dot S_z\big)=\big|\dot S_z\big)$, 
which simplifies Eq.~(\ref{Self.Eq}) to    
\begin{eqnarray}\label{Sigma.Eq}
\Sigma(s)&=&\frac{\beta}{\chi_0}\big(\dot S_z\big|\frac{i}{s+\mc{L}}\big|\dot S_z\big) 
=\frac{\beta}{\chi_0}\big(\dot S_z\big|\dot S_z\big)_s ~.
\end{eqnarray}
Hence, $\Sigma(\omega+i 0^+)$ is proportional to the Mori correlation function
$\big(\dot S_z \big| \dot S_z\big)_{\omega+i 0^+}$ and thus embodies the fluctuations 
of the ``flux" operator~\footnote{Correlation functions of this type of observables are called 
generalized Onsager-Casimir coefficients. They play a central role in the entropy production 
in irreversible processes~\cite{Fick,Zubarev,Kubo}.}
\begin{equation}\label{dotSz.Eq}
\dot S_z = \frac{1}{i\hbar}\big[S_z, \mc{V}\big] = \sum_\mu \mc{F}^\mu\otimes\mc{B}^\mu~,
\end{equation}
where $\mc{F}^\mu \equiv \frac{1}{i\hbar} \big[ S_z ,\mc{A}^\mu \big]$. For later convenience let us 
write the matrix elements of these operators: 
$\mc{F}^\mu_{ll'}\equiv\langle l|\mc{F}^\mu|l'\rangle= \frac{i}{\hbar}~\delta M_{l'l}~\mc{A}^\mu_{ll'}$, 
where $\delta M_{l'l}\equiv M_{l'}-M_l$.

We now employ a perturbation expansion for the evaluation of the self-energy $\Sigma(\omega+i 0^+)$. 
To this end, one needs to expand the resolvent $\mc{G}=i (s+\mc{L})^{-1}$ of Eq.~\ref{Sigma.Eq} 
in powers of $\mc{L}_\mc{V}$. Interestingly, $\dot S_z$ is linear in $\mc{V}$ and thus we may already 
obtain the lowest order effect by simply replacing $\mc{L}$ by $\mc{L}_0=\mc{L}_s+\mc{L}_{ph}$ in 
the denominator of Eq.~(\ref{MsSelf.Eq}). To second order then, 
$\big(\dot S_z \big| \dot S_z\big)_{\omega+i 0^+}$ equals to
\begin{eqnarray}\label{Lomega1.Eq}
\big(\dot S_z \big|\frac{i}{\mc{L}_0+\omega+i 0^+}\big|\dot S_z\big)=
\big(\dot S_z\big|\pi\delta\big(\mc{L}_0+\omega\big) \big|\dot S_z\big)\nonumber\\
=\pi\frac{e^{\beta\hbar\omega}-1}{\beta\hbar\omega}\big\langle 
\dot S_z\delta\big(\mc{L}_0+\omega\big)\dot S_z\big\rangle~,
\end{eqnarray}  
where we have used the reality of $\Sigma(\omega+i 0^+)$. Employing the representation 
$\delta(\mc{L}_0+\omega)=(2\pi)^{-1} \int dt~e^{i\mc{L}_0t}$, 
factorizing $e^{i\mc{L}_0t}=e^{i\mc{L}_st} e^{i\mc{L}_{ph}t}$, and using Eq.~(\ref{rho2.Eq}) we obtain   
\be\label{Lomega2.Eq}
\Sigma(\omega+i 0^+) =\frac{\beta}{\chi_0} \frac{e^{\beta\hbar\omega}-1}{2 \beta\hbar\omega} 
\sum_{\mu\mu'}\int d\omega' J^{s}_{\mu\mu'}(\omega') J^{ph}_{\mu\mu'}(\omega-\omega')~,  
\ee
where $J^{ph}_{\mu\mu'}(\omega)$ was defined previously, 
while $J^{s}_{\mu\mu'}(\omega)$ stand for the isolated spectral densities of the $\mc{F}^\mu$ operators, i.e.  
\be
J^{s}_{\mu\mu'}(\omega) = \int_{-\infty}^{\infty} dt~e^{i\omega t} 
\big\langle\mc{F}^\mu\mc{F}^{\mu'}(t)\big\rangle_{\text{s}}~,
\ee
where the thermal average is taken over the canonical ensemble $\rho_{\text{s}}(T)$.

\subsection{Markovian behavior and correlation frequency}\label{Pert.Subsec}
The origin of the mono-exponential decay of $S_z$ follows immediately from 
Eq.~(\ref{Lomega2.Eq}) and this is related to the discrete character of the spectrum. 
The spectral densities $J^s(\omega')$ of $\mc{F}_\mu$ are peaked at the characteristic Bohr frequencies 
$\omega'=(E_l-E_{l'})/\hbar$ of $\mc{H}_s$. On the other hand, we are interested in a 
long time regime: 
For all relevant experimental frequency scales (e.g. $\omega_L$ in NMR) 
$\omega \sim \omega_c \ll\omega'$, and thus, given the extremely short phonon correlation times, 
we may replace~\footnote{This holds true also for $\omega'=0$ given again the extremely short phonon correlation times. 
Besides, the frequency $\omega_B = 0$ is absent from the spectral decomposition of 
$\mc{J}^s$ since the corresponding matrix elements 
$\langle l| \mc{F}^s|l'\rangle \propto \langle l| \mc{A}^s|l'\rangle [M_{l'}-M_{l}]$  
vanish for $\hbar\omega_B=E_l-E_{l'}=0$.} $\mc{J}^{ph}(\omega-\omega')\simeq J^{ph}(-\omega')$,  
which is equivalent with the usual Markov approximation of replacing
$\Sigma(\omega + i 0^+) \simeq \Sigma(i 0^+)$ in Eq.~(\ref{MsSelf.Eq}). 
This ``memory-less'' character of the self-energy means that the fluctuations of the observables 
coupled to $S_z$ (i.e. the observables $\mc{F}^\mu$ that enter in $\dot{S}_z$) 
decay in times $\tau'_c$ much shorter than the decay time $\tau_c=1/\omega_c$ of $M(t)$. 
The physical origin of this separation of time scales can be attributed to the fact that, 
in contrast to $\mc{F}^\mu$, $S_z$ is a conserved variable for the isolated magnetic cluster 
and thus is expected to decay much slower in the presence of the small magnetoelastic coupling.   
At the same time, this separation of time scales is at the origin of the Markovian decay of $S_z$: 
As long as $t\gg \tau_c'$, one may replace $M(t-t')$ with $M(t)$ inside the integral of Eq.~(\ref{DiffEq.Eq}) 
and extend the upper limit to infinity, i.e.     
$\frac{d}{dt} M(t) \simeq - \int_0^\infty dt' \Sigma(t')\cdot M(t) \equiv -\Sigma(i 0^+)\cdot M(t)$, 
which gives\footnote{The same result can be obtained by an inverse Laplace transform of Eq.~(\ref{MsSelf.Eq}).}
\be\label{Mt2.Eq}
M(t)=(\chi_0/\beta)~e^{-\Sigma(i 0^+) t},~t\geq 0~.
\ee
We can thus identify the zero-frequency portion of the self-energy with the correlation frequency $\omega_c$:   
\begin{eqnarray}\label{omegac1.Eq}
\omega_c=\Sigma(i 0^+)=
\frac{\beta}{2\chi_0}\sum_{\mu\mu'}\int_{-\infty}^{\infty} 
d\omega'\mc{J}^{s}_{\mu\mu'}(\omega') \mc{J}^{ph}_{\mu\mu'}(-\omega').
\end{eqnarray}
On the other hand, according to Eqs.~(\ref{chi2.Eq}) and (\ref{fluctdiss.Eq}), 
the susceptibility and the spectral density have the familiar Debye (or Lorentzian) form:  
\begin{eqnarray}
\chi(s) &=& \chi_0 \frac{i \omega_c}{s + i \omega_c}\label{chi3.Eq}\\
\mc{J}(\omega)&=&\frac{2\hbar}{1-e^{-\beta\hbar\omega}}\chi_0\frac{\omega \omega_c}{\omega^2+
\omega_c^2}\label{Spectral2.Eq}~. 
\end{eqnarray}
Equations~(\ref{omegac1.Eq}), (\ref{chi3.Eq}) and (\ref{Spectral2.Eq}) are our central results. 
It is expedient to rewrite Eq.~(\ref{omegac1.Eq}) in a more convenient form, in terms of the various 
quantum-mechanical transition rates $W_{l l'}$ from $|l \rangle$ to $|l' \rangle$. 
To this end, we use the spectral representation 
$\mc{J}^{s}_{\mu\mu'}(\omega')= 
2\pi \sum_{l l'} P_l \mc{F}^\mu_{ll'} \mc{F}^{\mu'}_{l'l} \delta (\omega'-\omega_{ll'})$, 
where $P_l \equiv e^{-\beta E_l} / Z_s$ and $\mc{F}^{\mu}_{ll'}$ were given previously. We find
\be
\frac{\omega_c}{2\pi} = \frac{1}{2}\sum_{l,l'} P_l~\frac{\delta M_{l'l}^2}{\langle \delta S_z^2\rangle} 
\cdot \Big[ 
\frac{1}{\hbar^2}\sum_{\mu\mu'} \mc{A}^\mu_{ll'}\mc{A}^{\mu'}_{l'l} \mc{J}^{ph}_{\mu\mu'}(\omega_{l'l})
\Big]~.
\ee
The last factor of this equation can be identified to be the quantum-mechanical transition rate 
$W_{ll'}$ from $|l\rangle$ to $|l'\rangle$. We therefore obtain the convenient expression
\be\label{omegac2.Eq}
\frac{\omega_c}{2\pi} = \frac{1}{2} \sum_{l,l'} P_l~\frac{\delta M_{l'l}^2}{\langle \delta S_z^2\rangle}~
W_{ll'}~.
\ee
We remark here that we can eliminate the factor of $1/2$ appearing in the right hand sides 
of Eqs.~(\ref{omegac1.Eq}) and (\ref{omegac2.Eq}) by restricting the integral of (\ref{omegac1.Eq}) 
to positive (or negative) frequencies only, or by constraining the sum of (\ref{omegac2.Eq}) to run over 
$l<l'$. This is allowed by the detailed balance condition  
$\mc{J}_{\mu\mu'}^i(-\omega')=e^{-\beta\hbar\omega'}\mc{J}^i_{\mu'\mu}(\omega')$ ($i=$ s, ph) or
its equivalent form $P_l W_{ll'}=P_{l'} W_{l'l}$.

Let us now describe in qualitative terms the main features of Eq.~(\ref{omegac1.Eq}) or 
Eq.~(\ref{omegac2.Eq}). According to these expressions, $\omega_c$ 
is a weighted sum over all relevant resonance channels between the spin and the phonon system. 
Each contribution enters in a transparent and physically appealing 
factorized form. Energy conservation is ensured by the opposite frequency arguments in $\mc{J}^{s}$ 
and $\mc{J}^{ph}$. The one-phonon portion of the latter is proportional to the number of available 
(i.e. thermally excited) phonons in resonance with the electronic spin system. 
We also note that it is $\mc{F}_\mu$ and $\mc{B}_\mu$ triggering the decay of $S_z$ 
(cf. (\ref{dotSz.Eq})) that enter in Eq.~(\ref{omegac1.Eq}). This is also reflected in Eq.~(\ref{omegac2.Eq}) 
by the fact that transitions between two levels $l,l'$ contribute to $\omega_c$ only if they differ in their 
magnetic moment. 
More generally, each of the three known mechanisms of the slowing down effect can be now associated 
to one of the three major quantities appearing in Eq.~(\ref{omegac1.Eq}), 
namely $\mc{J}^{ph}$, $\mc{J}^{s}$ and $\chi_0$. In AFMR's (see also Sec.~\ref{NMR.Sec} below), 
it is the drastic reduction of the resonant portions of the phonon spectral weight 
$\mc{J}^{ph}$ as $T$ drops below the spin gap. 
In SMM's on the other hand, the effect appears already well above the typical intra-barrier 
excitations, and thus must be attributed to selection rules in $\mc{J}^s$ which necessitate an  
``over the barrier'' relaxation process. Finally, the appearance of the susceptibility in 
the denominator of Eq.~(\ref{omegac1.Eq}) is underlying the critical slowing down observed 
in SCM's~\cite{SCM1,SCM2,SCM3,SCM4,SCM5} similarly to bulk ferromagnets~\cite{Forster}.

We should finally note here that one may start with the magnetic energy $\mc{H}_s$ instead of $S_z$ 
and show, following exactly the same steps as above, that this observable decays also independently 
from the remaining slow degrees of freedom. Its corresponding decay rate $\omega_E$ can be also derived 
and the result is fully analogous to Eq.~(\ref{omegac2.Eq}):
\be\label{omegaHs.Eq}
\frac{\omega_E}{2\pi} = \frac{1}{2} \sum_{l,l'} P_l 
\frac{\delta E_{l'l}^2}{\langle \delta \mc{H}_s^2\rangle} W_{ll'}~,
\ee
where $\delta E_{l'l}\equiv E_{l'}-E_l$, 
$\langle \delta \mc{H}_s^2\rangle = k_B T^2 C_m $, and $C_m$ is the magnetic specific heat.
The ``magnetic energy correlation frequency'' $\omega_E$ is of experimental relevance in time-dependent 
specific heat measurements\cite{Mettes} and is distinct from $\omega_c$. 
It is nevertheless expected, based on the above discussion, that $\omega_E$ also manifests a dramatic 
slowing down similar to $\omega_c$.

\section{Nuclear spin-lattice relaxation rate in AFMR's}\label{NMR.Sec}
Here we revisit the problem of proton $1/T_1$ in AFMR's and clarify a 
number of aspects in the light of Eqs.~(\ref{Spectral2.Eq}) and (\ref{omegac1.Eq}) or (\ref{omegac2.Eq}). 
Quite generally, since the proton spin is a local probe, $1/T_1$ samples 
all (auto- and pair-) electronic spin spectral densities evaluated at $\omega_L$. 
Remarkably however, in clusters with equivalent ionic spins (such as AFMR's)
symmetry arguments together with the fact that $\omega_L$ is much smaller than the typical 
Bohr frequencies $\omega_B$ assert that away from level crossings the relevant 
terms contributing to $1/T_1$ are all proportional to the spectral density of 
the total magnetization~\cite{mythesis,NMR_Santini}, i.e. 
\be
\frac{1}{T_1}=A \mc{J}(\omega_L)~,
\ee 
where the proportionality constant $A$ embodies the specific details regarding
the geometry and the strength of the nuclear-electron hyperfine coupling~\cite{NMR_Borsa}. 
The remaining terms of the electronic spin spectral densities are vanishingly small 
and become relevant only at field-induced level crossings~\cite{mythesis}. 
For symmetric clusters then and using Eq.~(\ref{Spectral2.Eq}) 
(with $\beta\hbar\omega_L\ll 1$) we obtain
\begin{equation}\label{T1.Eq}
\frac{1}{T_1} =(2 k_B A) \chi_0 T \frac{\omega_c(T)}{\omega_L^2+\omega_c(T)^2}~,
\end{equation}
which is the Lorentzian form used previously~\cite{NMR_Baek,NMR_Santini}.

In AFMR's, $\mc{J}^{s}(\omega_B)/\big(\chi_0 T\big)$ does not change drastically
over the whole temperature range of interest. Hence, the observed dramatic decrease of $\omega_c$
and the resulting enhancement~\cite{NMR_Baek,NMR_Borsa,NMR_Santini} of $1/T_1$ when $\omega_c(T)=\omega_L$ 
(cf. Eq.~(\ref{T1.Eq})) stems, as mentioned above, from a drastic reduction of 
the number of available (thermally excited) phonons which are in resonance with the magnetic cluster. 
This is because on decreasing temperature the relevant low-lying magnetic excitations 
are increasingly gapped while the frequency peak of $\mc{J}^{ph}(\omega)$ shifts toward lower $\omega$.
Eventually the phonon spectral density peak shifts below $\Delta/\hbar$ 
and one-phonon processes become ineffective in driving the electronic spin fluctuations, thus resulting in a
drastic drop of $\omega_c$. 
One can also explain the observed field independence of $\omega_c$\cite{NMR_Baek,NMR_Santini,NMR_Borsa} 
at intermediate temperatures and for $B\ll\Delta/(g\mu_B)$ as follows. 
The relevant Bohr frequencies (with $\delta M_{l'l}\neq 0$) 
in these temperature and field regimes are $\Delta/\hbar\pm\omega_e$
$\omega_e$ and $2 \omega_e$ (here $\omega_e$ is the electron Larmor frequency). 
These frequencies correspond respectively to the transitions between the lowest singlet 
and the first triplet excitation (with $\delta M_{l'l}=\pm\hbar$) and to transitions within 
the lowest triplet. The later do not contribute appreciably since 
$\mc{J}^{ph}(\omega_e)\ll \mc{J}^{ph}(\Delta/\hbar)$. At the same time 
$\mc{J}^{ph}(\Delta/\hbar\pm\omega_e)\simeq \mc{J}^{ph}(\Delta/\hbar)$ 
and $\chi_0(B)\simeq\chi_0(0)$. As a result $\omega_c$ is field-independent at intermediate 
temperatures and $B \ll \Delta/(g\mu_B)$. A field dependence should only arise at higher fields and  
very low temperatures or for clusters with small $\Delta$.

\section{Summary}\label{Summary.Sec}
We have presented a microscopic theory for the omnipresent mono-exponential slowing down effect 
observed in a wide range of molecular nanomagnets. 
To this end, we first expressed the magnetization fluctuations in terms of the associated self-energy 
or memory function which embodies the fluctuations of the observables that couple to $S_z$. 
It is then shown that the discreteness of the magnetic energy spectrum guarantees that the self-energy 
decays in times much shorter than $1/\omega_c$. This separation of time scales results in a dynamical 
decoupling of the magnetization fluctuations and thus explains their observed Markovian behavior.
In addition, we have derived an analytical expression for the correlation frequency parameter $\omega_c$
which embodies in a convenient factorized form all essential ingredients. 
This expression highlights the three specific mechanisms of the slowing down effect which are known so far 
in nanomagnets, namely (i) the existence of an anisotropy barrier, (ii) the increasingly gapped character
of the magnetic energy spectrum at low energies, and (iii) the possibility of a critical 
slowing down. In addition, this formula contains, in a convenient factorized form, the noise spectra 
of the relevant low-energy strain fields which are at present largely unexplored. 
Hence this work paves the way for a more direct and quantitative comparison with experiment 
and a deeper understanding of the underlying microscopic relaxation channels present 
in all nanomagnetic systems.

\section{Acknowledgments}
I would like to thank M.~Luban, A.~L\"auchli, K.~P.~Schmidt, M.~Belesi and F.~Mila for very fruitful 
discussions and valuable comments on the manuscript.

\end{document}